\documentclass[a4paper,oneside,11pt]{article} 

\usepackage[latin1]{inputenc}
\usepackage[T1]{fontenc}     
\usepackage{amsmath, amsthm} 
\usepackage{amsfonts,amssymb}
\usepackage{fullpage}

\usepackage{graphicx}        
\usepackage{subfig}          
\usepackage{wrapfig} 

\usepackage{array}           
\usepackage{multirow}        

\usepackage{listings}        
\usepackage{color}           
\usepackage{fancybox}        
\usepackage{bbm}             
\usepackage{dsfont}          
\usepackage{fancyhdr}        
\usepackage{natbib}

\usepackage{hyperref}
 \hypersetup{colorlinks, linkcolor=black, urlcolor=black} 

\makeatletter
\newif\if@restonecol
\makeatother

\usepackage{algorithm2e}
\usepackage{multirow}

\newcommand{\OMIT}[1]{}

\newcommand{\RR}{\mathbb{R}}

\begin{document}

\title{The influence of feature selection methods on accuracy, stability and interpretability of molecular signatures}

\author{
\small\bf Anne-Claire Haury\thanks{To whom correspondance should
    be addressed: \small 35, rue Saint Honor\'e, F-77300 Fontainebleau, France.}\\
\small Mines ParisTech, CBIO\\
\small Institut Curie, Paris, F-75248\\
\small INSERM, U900, Paris, F-75248\\
\small \texttt{anne-claire.haury@ensmp.fr}
\and
\small\bf Pierre Gestraud\\
\small Mines ParisTech, CBIO\\
\small Institut Curie, Paris, F-75248\\
\small INSERM, U900, Paris, F-75248\\
\small \texttt{pierre.gestraud@curie.fr} \\
\and
\small\bf Jean-Philippe Vert \\
\small Mines ParisTech, CBIO\\
\small Institut Curie, Paris, F-75248\\
\small INSERM, U900, Paris, F-75248\\
\small \texttt{jean-philippe.vert@mines-paristech.fr} \\
}


\maketitle

\begin{abstract}
\textbf{Motivation} Biomarker discovery from high-dimensional data is a crucial problem with enormous applications in biology and medicine. It is also extremely challenging from a statistical viewpoint, but surprisingly few studies have investigated the relative strengths and weaknesses of the plethora of existing feature selection methods.

\textbf{Methods} We compare $32$ feature selection methods on $4$ public gene expression datasets for breast cancer prognosis, in terms of predictive performance, stability and functional interpretability of the signatures they produce. 

\textbf{Results} We observe that the feature selection method has a significant influence on the accuracy, stability and interpretability of signatures. Simple filter methods generally outperform more complex embedded or wrapper methods, and ensemble feature selection has generally no positive effect. Overall a simple Student's t-test seems to provide the best results. 

\textbf{Availability} Code and data are publicly available at \\
http://cbio.ensmp.fr/$\sim$ahaury/.
\end{abstract}

\section{Introduction}
Biomarker discovery from high-dimensional data, such as transcriptomic or SNP profiles, is a crucial problem with enormous applications in biology and medicine, such as diagnosis, prognosis, patient stratification in clinical trials or prediction of the response to a given treatment. Numerous studies have for example investigated so-called \emph{molecular signatures}, i.e., predictive models based on the expression of a small number of genes, for the stratification of early breast cancer patients into low-risk or high-risk of relapse, in order to guide the need for adjuvant therapy \citep{Sotiriou2009Gene-Expression}.

While predictive models could be based on the expression of more than a few tens of genes, several reasons motivate the search for short lists of predictive genes. First, from a statistical and machine learning perspective, restricting the number of variables is often a way to reduce over-fitting when we learn in high dimension from few samples and can thus lead to better predictions on new samples. Second, from a biological viewpoint, inspecting the genes selected in the signature may shed light on biological processes involved in the disease and suggest novel targets. Third, and to a lesser extent, a small list of predictive genes allows the design of cheap dedicated prognostic chips.

Published signatures share, however, very few genes in common, raising questions about their biological significance \citep{Ioannidis2005Microarrays}. Independently of differences in cohorts or technologies, \citet{Ein-Dor2005Outcome} and \citet{Michiels2005Prediction} demonstrate that a major cause for the lack of overlap between signatures is that many different signatures lead to similar predictive accuracies, and that the process of estimating a signature is very sensitive to the samples used in the phase of gene selection. Specifically, \citet{Ein-Dor2006Thousands} suggest that many more samples than currently available would be required to reach a descent level of signature stability, meaning in particular that no biological insight should be expected from the analysis of current signatures. On the positive side, some authors noticed that the biological functions captured by different signatures  are similar, in spite of the little overlap between them at the gene level \citep{shen2008pathway,reyal2008comprehensive,wirapati2008meta}. 

From a machine learning point of view, estimating a signature from a set of expression data is a problem of \emph{feature selection}, an active field of research in particular in the high-dimensional setting \citep{Guyon2003introduction}. While the limits of some basic methods for feature selection have been highlighted in the context of molecular signatures, such as gene selection by Pearson correlation with the output \citep{Ein-Dor2006Thousands}, there are surprisingly very few and only partial investigations that focus on the \emph{influence of the feature selection method} on the performance and stability of the signature. \citet{lai2006comparison} compared various feature selection methods in terms of predictive performance only, and \citet{Abeel2010Robust} suggest that ensemble feature selection improves both stability and accuracy of SVM recursive feature elimination (RFE), without comparing it with other methods. However, it remains largely unclear how ''modern'' feature selection methods such as the elastic net \citep{Zou2005Regularization}, SVM RFE or stability selection \citep{Meinshausen2010Stability} behave in these regards and how they compare to more basic univariate techniques.

Here we propose an empirical comparison of a panel of feature selection techniques in terms of accuracy and stability, both at the gene and at the functional level. Using four breast cancer datasets, we observe significant differences between the methods. Surprisingly, we find that ensemble feature selection, i.e., combining multiple signatures estimated on random subsamples, has generally no positive impact, and that simple filters can outperform more complex wrapper or embedded methods. 

\section{Methods}

\subsection{Feature selection methods}
We compare eight common feature selection methods to estimate molecular signatures. All methods take as input a matrix of gene expression data for a set of samples from two categories (good and bad prognosis in our case), and return a set of genes of a user-defined size $s$. These genes can then be used to estimate a classifier to predict the class of any sample from the expression values of these genes only. Feature selection methods are usually classified into three categories \citep{Kohavi1997Wrappers,Guyon2003introduction}: \emph{filter methods} select subsets of variables as a pre-processing step, independently of the chosen predictor; \emph{wrapper methods} utilize the learning machine of interest as a black box to score subsets of variable according to their predictive power; finally, \emph{embedded methods} perform variable selection in the process of training and are usually specific to given learning machines. We have selected popular methods representing these three classes, as described below.

\subsubsection{Filter methods}

Univariate filter methods rank all variables in terms of relevance, as measured by a score which depends on the method. They are simple to implement and fast to run. To obtain a signature of size $s$, one simply takes the top $s$ genes according to the score. We consider the following four scoring functions to rank the genes: the \emph{Student's t-test} and \emph{Wilcoxon sum-rank test}, which evaluate if each feature is differentially expressed between the two classes; and the \emph{Bhattacharyya distance} and \emph{relative entropy} to calculate a distance between the distributions of the two groups. We used the MATLAB Bioinformatics toolbox to compute these scoring functions.

\subsubsection{Wrapper methods}
Wrapper methods attempt to select jointly sets of variables with good predictive power for a predictor. Since testing all combinations or variables is computationally impossible, wrapper methods usually perform a greedy search in the space of sets of features. We test \emph{SVM recursive feature elimination (RFE)} \citep{Guyon2002Gene}, which starts with all variables and iteratively removes the variables which contribute least to a linear SVM classifier trained on the current set of variables. We remove $20\%$ of features at each iteration until $s$ remain, and then remove them one by one in order to rigourously rank the first $s$. Following \citep{Abeel2010Robust}, we set the SVM parameter $C$ to $1$, and checked afterwards that other values of $C$ did not have a significant influence on the results. Alternatively, we test a \emph{Greedy Forward Selection (GFS)} strategy for least squares regression also termed Orthogonal Matching Pursuit, where we start from no variable and add them one by one by selecting each time the one which minimizes the sum of squares, in a $3$-fold internal cross-validation setting. This algorithm was implemented in the SPAMS toolbox for Matlab initially published along with \citet{mairal2010online}.

\subsubsection{Embedded methods}

Embedded methods are learning algorithms which perform feature selection in the process of training. We test the popular \emph{Lasso} regression \citep{Tibshirani1996Regression}, where a sparse linear predictor $\beta \in \RR^p$ is estimated by minimizing the objective function $R(\beta)+\lambda \|\beta\|_1$, where $R(\beta)$ is the mean square error on the training set (considering the two categories as $\pm 1$ values) and $\|\beta\|_1 = \sum_{i=1}^p|\beta_i|$. $\lambda$ controls the degree of sparsity of the solution, i.e., the number of features selected. We fix $\lambda$ as the smallest value which gives a signature of the desired size $s$. Alternatively, we tested the elastic net \citep{Zou2005Regularization}, which is similar to the Lasso but where we replace the $\ell_1$ norm of $\beta$ by a combination of the $\ell_1$ and $\ell_2$ norms, i.e., we minimize $R(\beta) +\lambda \|\beta\|_1 + \lambda/2 \|\beta\|_2^2$ and $\|\beta\|_2^2 = \sum_{i=1}^p\beta_i^2$. By allowing the selection of correlated predictive variables, the elastic net is supposed to be more robust than the Lasso while still selecting predictive variables. Again, we tune $\lambda$ to achieve a user-defined level of sparsity. For both algorithms, we used the code implemented in the SPAMS toolbox.

\subsection{Ensemble feature selection}

Many feature selection methods are known to be sensitive to small perturbations of the training data, resulting in unstable signatures. In order to ''stabilize'' variable selection, several authors have proposed to use ensemble feature selection on bootstrap samples: the variable selection method is run on several random subsamples of the training data, and the different lists of variables selected are merged into a hopefully more stable subset \citep{Bi2003Dimensionality, Meinshausen2010Stability, Abeel2010Robust}.

For each feature selection method described above, we tested in addition the following three aggregation strategies for ensemble feature selection. We first bootstrap the training samples $B=50$ times (i.e., draw a sample of size $n$ from the data with replacement $B$ times) to get $B$ rankings $(r^1...r^B)$ of all features by applying the feature selection method on each sample. For filter methods, the ranking of features is naturally obtained by decreasing score. For RFE and GFS, the ranking is the order in which the features are added or removed in the iterative process. For Lasso and elastic net, the ranking is the order in which the variables become selected when $\lambda$ decreases. We then aggregate the $B$ lists by computing a score $S_j = 1/B \sum_{b=1}^B f(r_j^b)$ for each gene $j$ as an average function of its rank $r_j^b$ in the $b$-th bootstrap experiment. We test the following functions of the rank for aggregation:
\begin{itemize}
\item \emph{Ensemble-mean \citep{Abeel2010Robust}}: we simply average the rank of a gene over the bootstrap experiments, i.e., we take $f(r) = r$ .
\item \emph{Ensemble-stability selection \citep{Meinshausen2010Stability}}: we measure the percentage of bootstrap samples fro which the gene ranks in the top $s$, i.e., $f(r) = 1$ if $r\leq s$, $0$ otherwise.
\item \emph{Ensemble-exponential}: we propose a soft version of stability selection, where we average an exponentially decreasing function of the rank, namely $f(r) = \exp{-r/s}$.
\end{itemize}
Finally, for each rank aggregation strategy, the aggregated list is the set of $s$ genes with the largest score.

\subsection{Accuracy of a signature}\label{sec:method_acc}
In order to measure the predictive accuracy of a feature selection method, we assess the performance of various supervised classification algorithms trained on the data restricted to the selected signature. More precisely, we tested $5$ classification algorithms: nearest centroids (NC), k-nearest neighbors (KNN) with $k=9$, linear SVM with $C=1$, linear discriminant analysis (LDA) and naive Bayes (BAYES). The parameters of the KNN and SVM methods were fixed to arbitrary default values, and we checked that no significantly better results could be obtained with other parameters by testing a few other parameters. We assess the performance of a classifier by the area under the ROC curve (AUC), in two different settings. First, on each dataset, we perform a 10-fold cross-validation (CV) experiment, where both feature selection and training of the classifier are performed on $90\%$ of the data, and the AUC is computed on the remaining $10\%$ of the data. This is a classical way to assess the relevance of feature selection of a given dataset. Second, to assess the performance of the signature across datasets, we estimate a signature on one dataset, and assess its accuracy on other datasets by again running a 10-fold CV experiment where only the classifier (restricted to the genes in the signature) is retrained on each training set. In both cases, we report the mean AUC across the folds and datasets, and assess the significance of differences between methods with a two-sided paired t-test.

\subsection{Stability of a signature}\label{sec:stab}
To assess the stability of feature selection methods, we compare signatures estimated on different samples in various settings. First, to evaluate stability with respect to small perturbation of the training set, we randomly subsample each dataset into pairs of subsets with $80\%$ of sample overlap, estimate a signature on each subset, and compute the overlap between two signatures in a pair as the fraction of shared genes, i.e., $|S_1\cap S_2|/s$. Note that this corresponds to the figure of merit defined by \citet{Ein-Dor2006Thousands}. The random sampling of subsets is repeated $20$ times on each dataset, and the stability values are averaged over all samples. We will refer to this procedure the \emph{soft-perturbation} setting in the remaining. Second, to assess stability with respect to strong perturbation within a dataset, we repeat the same procedure but this time with no overlap between two subsets of samples. In practice, we can only sample subsets of size $N/2$, where $N$ is the number of samples in a dataset, to ensure that they have no overlap. Again, we measure the overlap between the signatures estimated on training sets with no sample in common. We call this procedure the \emph{hard-perturbation} setting. Finally, to assess the stability across datasets, we estimate signatures on each dataset independently, using all samples on each dataset, and measure their overlap. We call this procedure the \emph{between-datasets setting} below.

\subsection{Functional interpretability and stability of a signature}\label{sec:GO}
To interpret a signature in terms of biological functions, we perform functional enrichment analysis by inspecting the signature for over-represented Gene Ontology (GO) terms. This may hint at biological hypothesis underlying the classification \citep{shen2008pathway,reyal2008comprehensive}. We performed a hypergeometrical test on each of the $5830$ GO biological process (BP) terms that were associated to at least one gene in our dataset, and corrected the resulting p-values for multiple testing through the procedure of \citet{benjamini1995controlling}. To assess the \emph{interpretability} of a signature, i.e., how easily one can extract a biological interpretation, we computed the number of GO terms over-represented at $5\%$ FDR. To compare two signatures in functional terms, we first extracted from each signature the list of 10 GO terms with the smallest p-values, and compared the two lists of GO terms by the similarity measure of \citet{wang2007new} which takes into account not only the overlap between the lists but also the relationships between GO BP. Finally, to assess the \emph{functional stability} of a selection method, we followed a procedure similar to the one presented in Section \ref{sec:stab} and measured the mean functional similarity of signatures in the soft-perturbation, hard-perturbation and between-datasets settings.

\section{Data}\label{sec:material}
We collected $4$ breast cancer datasets from Gene Expression Omnibus \citep{barrett2009ncbi}, as described in Table \ref{tab:data}. The four datasets address the same problem of predicting metastatic relapse in breast cancer on different cohorts, and were obtained with the Affymetrix HG-U133A technology. We used a custom CDF file with EntrezGene ids as identifiers \citep{dai2005evolving} to estimate expression levels for $12,065$ genes on each array, and normalized all arrays with the Robust Multi-array Average procedure \citep{Irizarry2003Exploration}.
\begin{table}[!ht]
\centering
\begin{tabular}{|l|l|l|l|l|} \hline
Dataset name     	& $\sharp$ examples & $\sharp$ positives & source \\ \hline
GSE1456 		& $159$             & $40$               &  \citet{pawitan2005gene}\\
GSE2034                  & $286$             & $107$ 		 &  \citet{Wang2005Gene-expression}  \\
GSE2990     	           &  $125$            & $49$               &  \citet{sotiriou2006gene} \\
GSE4922               &  $249$            & $89$               &  \citet{ivshina2006genetic}\\\hline
\end{tabular}
\\
\caption{The four breast cancer datasets used in this study.}
\label{tab:data}
\end{table}

\section{Results}
\subsection{Accuracy}

We first assess the accuracy of signatures obtained by different feature selection methods. Intuitively, the accuracy refers to the performance that a classifier trained on the genes in the signature can reach in prediction. Although some feature selection methods (wrapper and embedded) jointly estimate a predictor, we dissociate here the process of selecting a set of genes and training a predictor on these genes, in order to perform a fair comparison common to all feature selection methods. We tested the accuracy of 100-gene signatures obtained by each feature selection method, combined with 5 classifiers to build a predictor as explained in Section \ref{sec:method_acc}. Table \ref{tab:area_compare} shows the mean accuracies (in AUC) over the datasets reached by the different combinations in 10-fold cross-validation. 

\begin{figure}
\centering
\includegraphics[width=\textwidth]{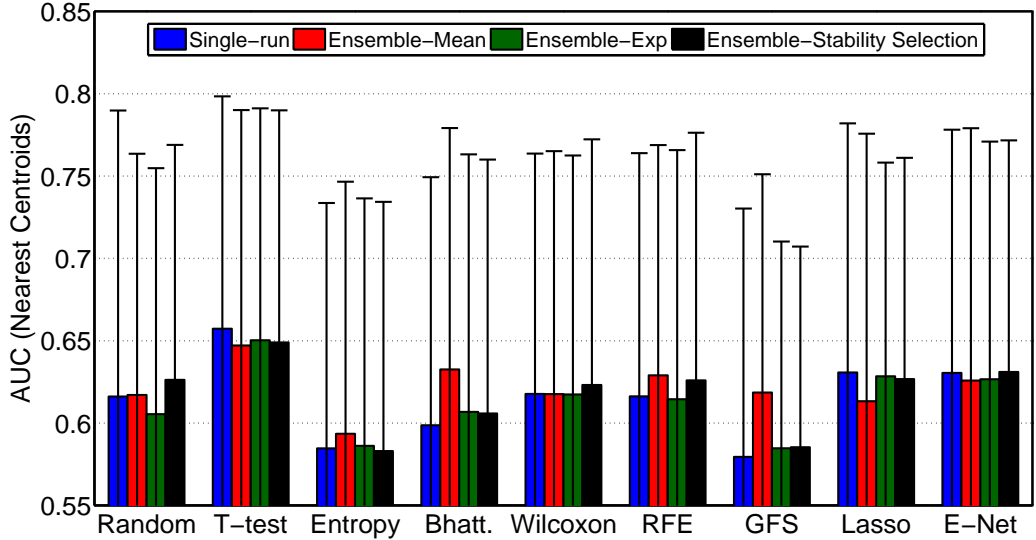}
\caption{Area under the ROC curve for a signature of size $100$ in a $10$-fold CV setting and averaged over the four datasets}
\label{fig:auc_NC}
\end{figure}

Globally, we observe only limited differences between the feature selection methods, for a given classification method. In particular the selection of a random signature reaches a baseline AUC comparable to that of other methods, confirming results already observed by \cite{Ein-Dor2005Outcome}. Second, we observe that, among all classification algorithms, the simple NC classifier consistently gives good results compared to other classifiers. We therefore choose it as a default classification algorithm for further assessment of the performance of the signatures below. Figure \ref{fig:auc_NC} depicts graphically the AUC reached by each feature selection method with NC as a classifier, reproducing the first three lines of Table \ref{tab:area_compare}. In the single-run framework, the t-test performs significantly better than most methods ($p<0.001$ against random, $p<0.01$ against entropy, Bhattacharyya, Wilcoxon and GFS). Lasso and Elastic Net perform similarly and show an AUC significantly higher than GFS and Entropy ($p<0.05$). Except for the t-test, random feature selection is not significantly worse than any other algorithm. Finally, we observe that ensemble methods for feature selection do not bring any improvement in accuracy in general since only Bhattacharyya and GFS benefit from ensemble-mean (resp. $p<0.05$ and $p<0.1$) and no significant improvement is obtained from the use of the two remaining ensemble aggregation methods. 

\begin{table*}[!ht] 
\begin{tiny} 
\begin{center} 
\begin{tabular}{|l|l|c|c|c|c|c|c|c|c|c|} 
\hline 
 Class.	&Type			& Random    & t-test 	&Entropy & Bhatt. & Wilcoxon & SVM RFE 	& GFS 	    & Lasso     & Elastic Net \\ \hline \hline 
\multirow{3}{*}{NC} & S& 0.62(0.17)& \textcolor{red}{0.66(0.14)}& 0.58(0.15)& 0.60(0.15)& 0.62(0.15)& 0.62(0.15)& 0.58(0.15)& 0.63(0.15)& 0.63(0.15)\\ 
 & E-M& 0.62(0.15)& 0.65(0.14)& \textcolor{red}{0.59(0.15)}& \textcolor{red}{0.63(0.15)}& 0.62(0.15)& \textcolor{red}{0.63(0.14)}& \textcolor{red}{0.62(0.13)}& 0.61(0.16)& 0.63(0.15)\\ 
 & E-E& 0.61(0.15)& 0.65(0.14)& 0.59(0.15)& 0.61(0.16)& 0.62(0.15)& 0.61(0.15)& 0.58(0.13)& 0.63(0.13)& 0.63(0.14)\\ 
 & E-S& \textcolor{red}{0.63(0.14)}& 0.65(0.14)& 0.58(0.15)& 0.61(0.15)& 0.62(0.15)& 0.63(0.15)& 0.59(0.12)& 0.63(0.13)& \textcolor{red}{0.63(0.14)}\\ \hline 
\multirow{3}{*}{KNN} & S& 0.59(0.16)& 0.61(0.15)& 0.52(0.11)& 0.57(0.13)& 0.63(0.15)& 0.60(0.15)& 0.59(0.13)& 0.60(0.17)& 0.60(0.17)\\ 
 & E-M& 0.61(0.14)& 0.62(0.15)& 0.57(0.15)& 0.60(0.15)& \textcolor{red}{0.64(0.16)}& 0.62(0.15)& 0.61(0.12)& 0.61(0.15)& 0.60(0.12)\\ 
 & E-E& 0.55(0.13)& 0.63(0.15)& 0.53(0.10)& 0.54(0.10)& 0.63(0.16)& 0.60(0.17)& 0.54(0.16)& 0.61(0.14)& 0.60(0.17)\\ 
 & E-S& 0.60(0.13)& 0.63(0.15)& 0.54(0.11)& 0.54(0.12)& 0.62(0.16)& 0.58(0.14)& 0.55(0.14)& 0.62(0.14)& 0.60(0.14)\\ \hline 
\multirow{3}{*}{LDA} & S& 0.54(0.12)& 0.56(0.12)& 0.51(0.14)& 0.55(0.13)& 0.52(0.12)& 0.56(0.12)& 0.50(0.13)& 0.58(0.14)& 0.57(0.14)\\ 
 & E-M& 0.53(0.10)& 0.55(0.13)& 0.55(0.13)& 0.58(0.12)& 0.56(0.13)& 0.60(0.15)& 0.52(0.14)& 0.59(0.14)& 0.60(0.13)\\ 
 & E-E& 0.54(0.13)& 0.53(0.15)& 0.52(0.15)& 0.53(0.11)& 0.53(0.14)& 0.57(0.13)& 0.53(0.15)& 0.59(0.12)& 0.58(0.13)\\ 
 & E-S& 0.54(0.13)& 0.52(0.13)& 0.54(0.13)& 0.55(0.12)& 0.52(0.14)& 0.57(0.16)& 0.54(0.15)& 0.59(0.15)& 0.60(0.13)\\ \hline 
\multirow{3}{*}{NB} & S& 0.57(0.14)& 0.60(0.13)& 0.58(0.11)& 0.58(0.14)& 0.57(0.13)& 0.56(0.14)& 0.54(0.11)& 0.59(0.15)& 0.59(0.15)\\ 
 & E-M& 0.59(0.13)& 0.59(0.14)& 0.57(0.14)& 0.59(0.13)& 0.57(0.13)& 0.56(0.13)& 0.59(0.12)& 0.57(0.15)& 0.57(0.14)\\ 
 & E-E& 0.55(0.15)& 0.60(0.14)& 0.58(0.12)& 0.57(0.13)& 0.58(0.13)& 0.57(0.14)& 0.58(0.11)& 0.58(0.12)& 0.58(0.13)\\ 
 & E-S& 0.58(0.14)& 0.60(0.14)& 0.57(0.13)& 0.57(0.13)& 0.58(0.13)& 0.56(0.14)& 0.58(0.10)& 0.58(0.11)& 0.58(0.13)\\ \hline 
\multirow{3}{*}{SVM} & S& 0.56(0.18)& 0.56(0.15)& 0.55(0.11)& 0.55(0.12)& 0.54(0.15)& 0.62(0.14)& 0.51(0.16)& 0.62(0.15)& 0.62(0.15)\\ 
 & E-M& 0.51(0.15)& 0.55(0.14)& 0.59(0.16)& 0.60(0.13)& 0.56(0.13)& 0.62(0.15)& 0.55(0.16)& 0.61(0.16)& 0.61(0.16)\\ 
 & E-E& 0.54(0.16)& 0.54(0.15)& 0.54(0.13)& 0.54(0.12)& 0.55(0.15)& 0.61(0.17)& 0.56(0.17)& \textcolor{red}{0.63(0.13)}& 0.62(0.16)\\ 
 & E-S& 0.54(0.17)& 0.55(0.18)& 0.56(0.12)& 0.56(0.12)& 0.54(0.14)& 0.61(0.16)& 0.55(0.17)& 0.63(0.14)& 0.62(0.16)\\ \hline 
\end{tabular}
\end{center} 
\caption{AUC obtained for each combination of feature selection and classification method, in 10-fold cross validation and averaged over the datasets. Standard error is shown within parentheses. For each selection algorithm, we highlighted the setting in which it obtained the best performance. The \emph{Type} column refers to the use of feature selection run a single time (S) or through ensemble feature selection, either with the mean (E-M), exponential (E-E) or stability selection (E-S) procedure to aggregate lists. } 
\label{tab:area_compare} 
\end{tiny} 
\end{table*}

\begin{figure*}[ht]
\centering
\includegraphics[width=\textwidth]{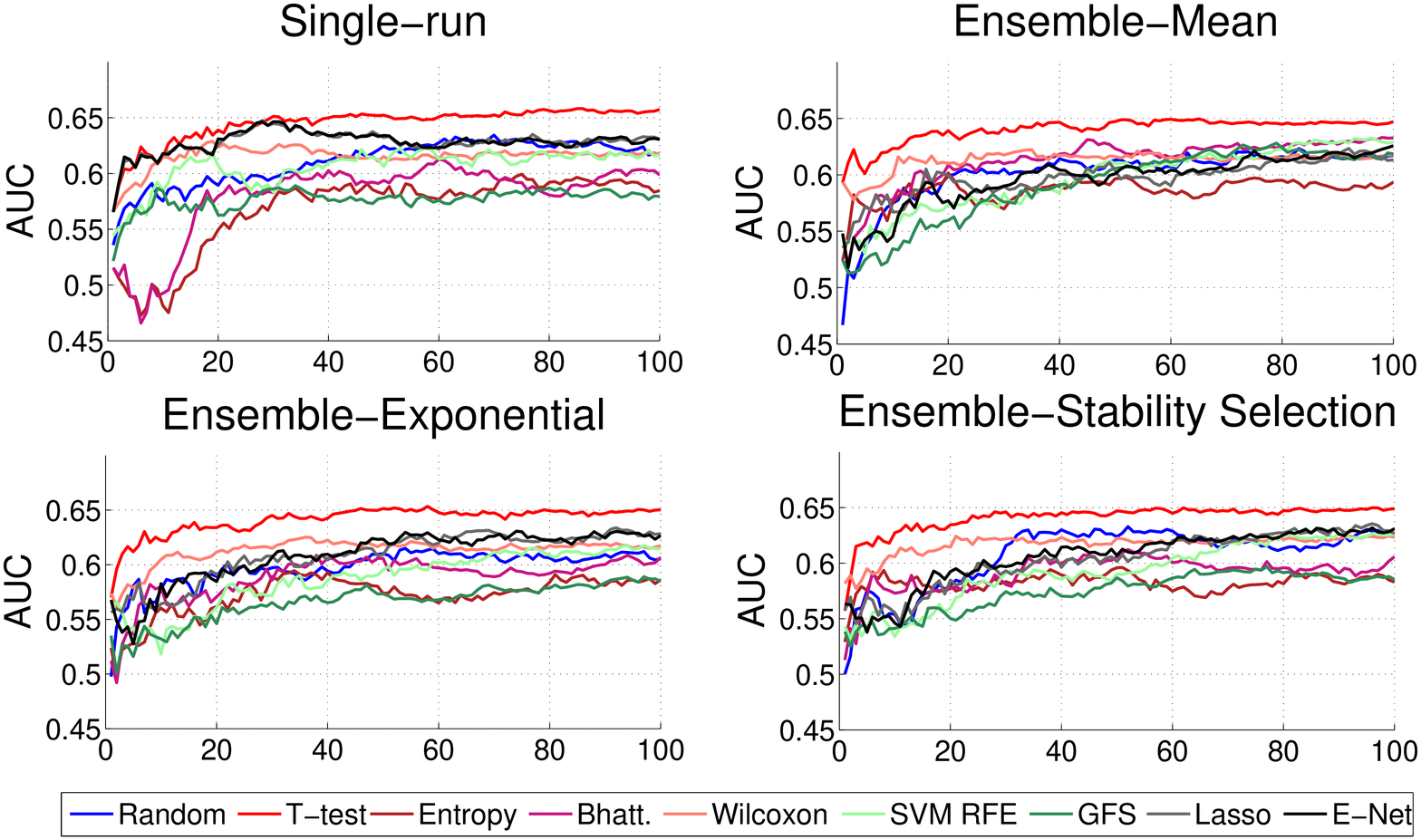}
\caption{AUC of a NC classifier trained as a function of the size of the signature, for different feature selection methods, in a $10$-fold CV setting averaged over the four datasets}
\label{fig:auc_size}
\end{figure*}

In order to check how these results depend on the size of the signature, we plot in Figure \ref{fig:auc_size} the AUC of the $9$ feature selection methods, with or without ensemble averaging, combined with a NC classifier, as a function of the size of the signature. Interestingly, we observe that in some cases the AUC seems to increase early, implying that fewer than $100$ genes may be sufficient to obtain the maximal performance. Indeed, while it is significant that $100$-gene signatures perform better than a list of fewer than $10$ features ($p<0.05$ regardless of the method or the setting), signatures of size $50$ do not lead to significantly worse performances in general. It is worth noting that some algorithms have an increasing AUC curve in this range of sizes, and we observe no overfitting that may lead to a decreasing AUC when the number of features increases. Random selection was previously shown to give an AUC equivalent to other methods for a large signature, but as we observe on this picture, the fewer genes the larger the gap in AUC. 

Finally, we estimate the predictive performance of a signature across datasets (Supplementary Table 1). Entropy is significantly less accurate than all other methods. T-test significantly outperforms other filter methods, and elastic net and Lasso also perform significantly better than Wilcoxon and SVM RFE. T-test and SVM RFE benefit from ensemble-mean, but no method significantly benefits from ensemble-exponential or ensemble-stability selection.

\subsection{Stability of gene lists}
\begin{figure}[!ht]
  \centering
  \subfloat[]{\label{fig:stab100}\includegraphics[width=.75\textwidth]{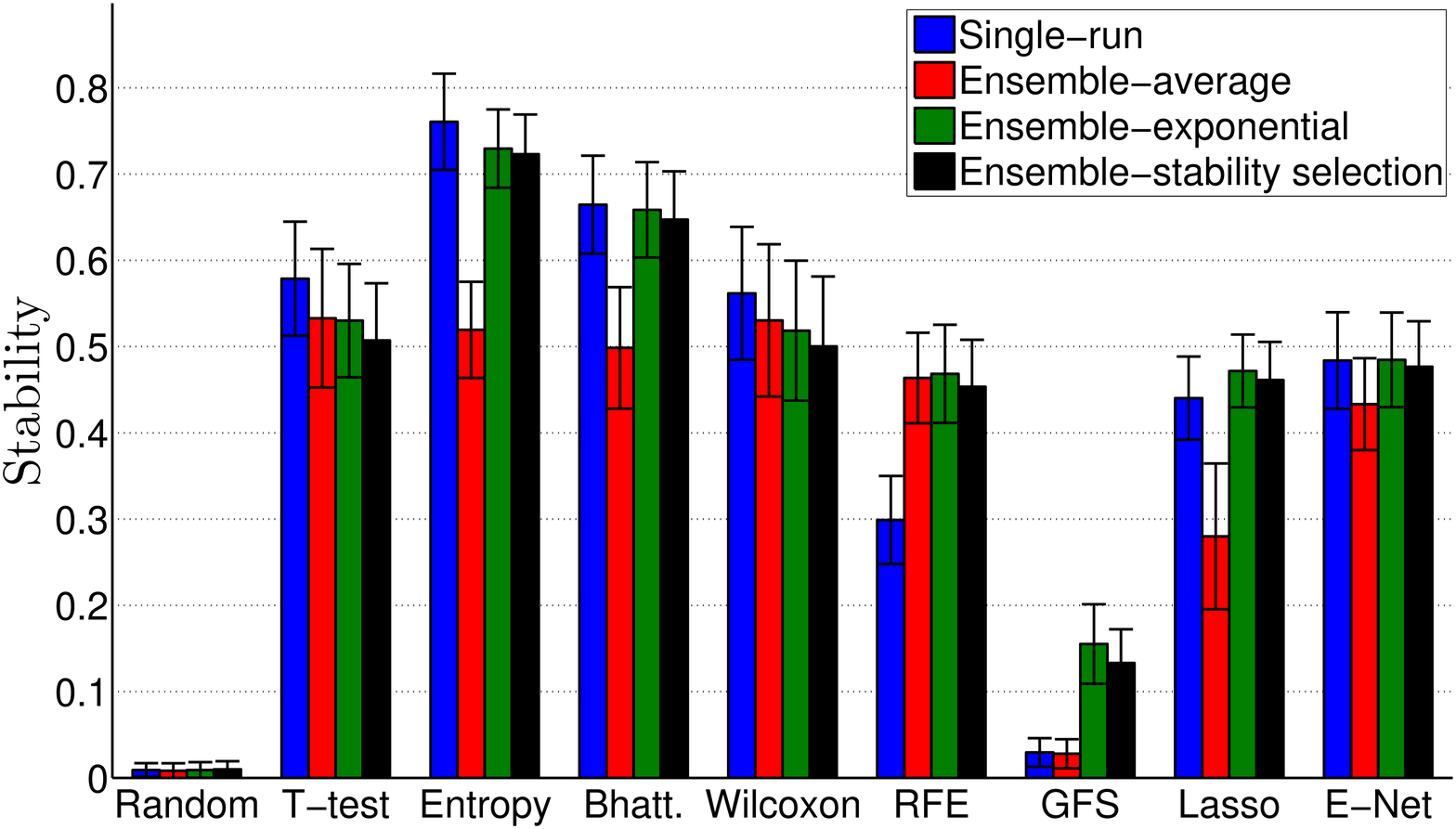}}

  \subfloat[]{\label{fig:stab100_hard}\includegraphics[width=.75\textwidth]{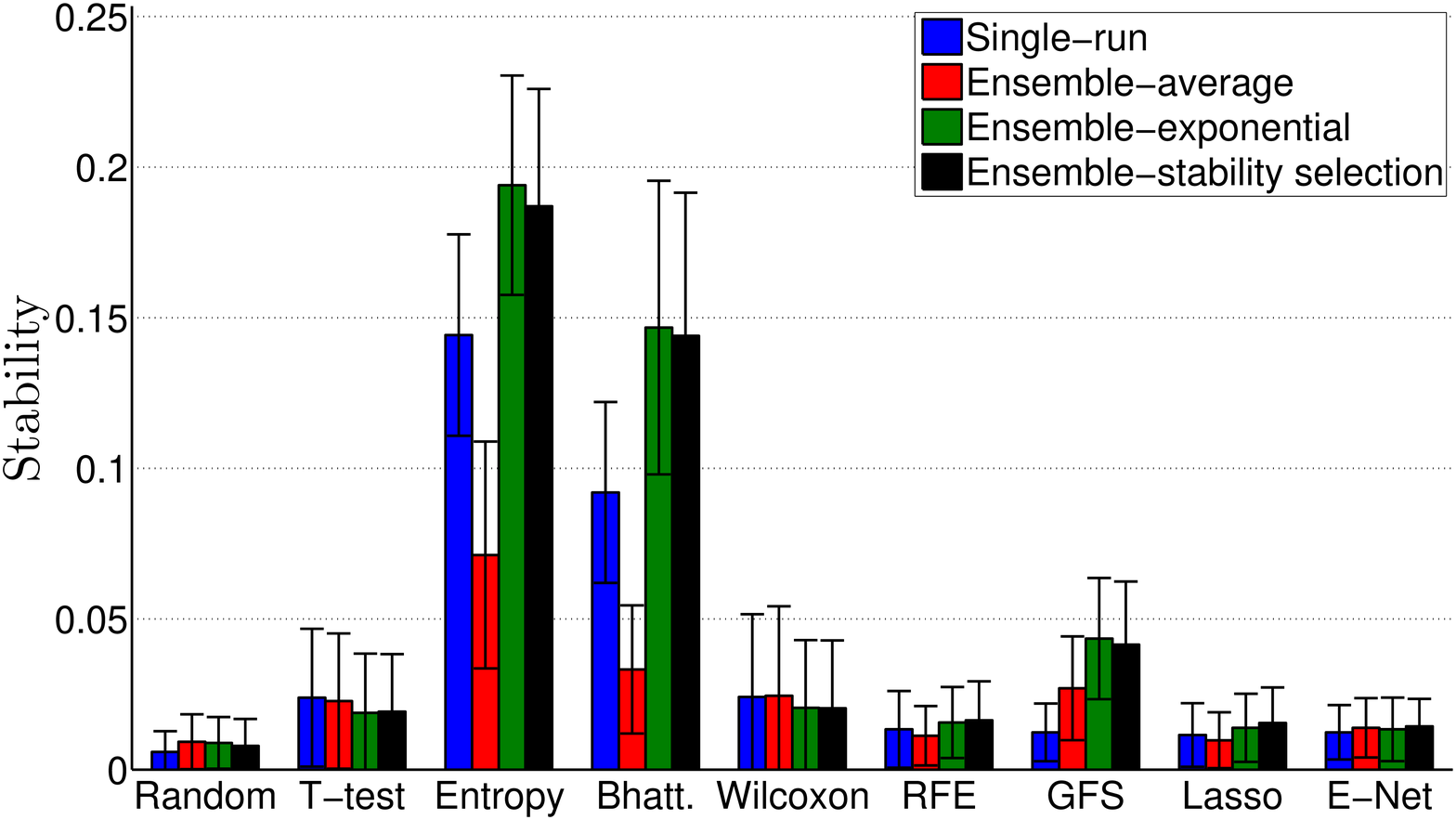}}

  \subfloat[]{\label{fig:stab_between}\includegraphics[width=.75\textwidth]{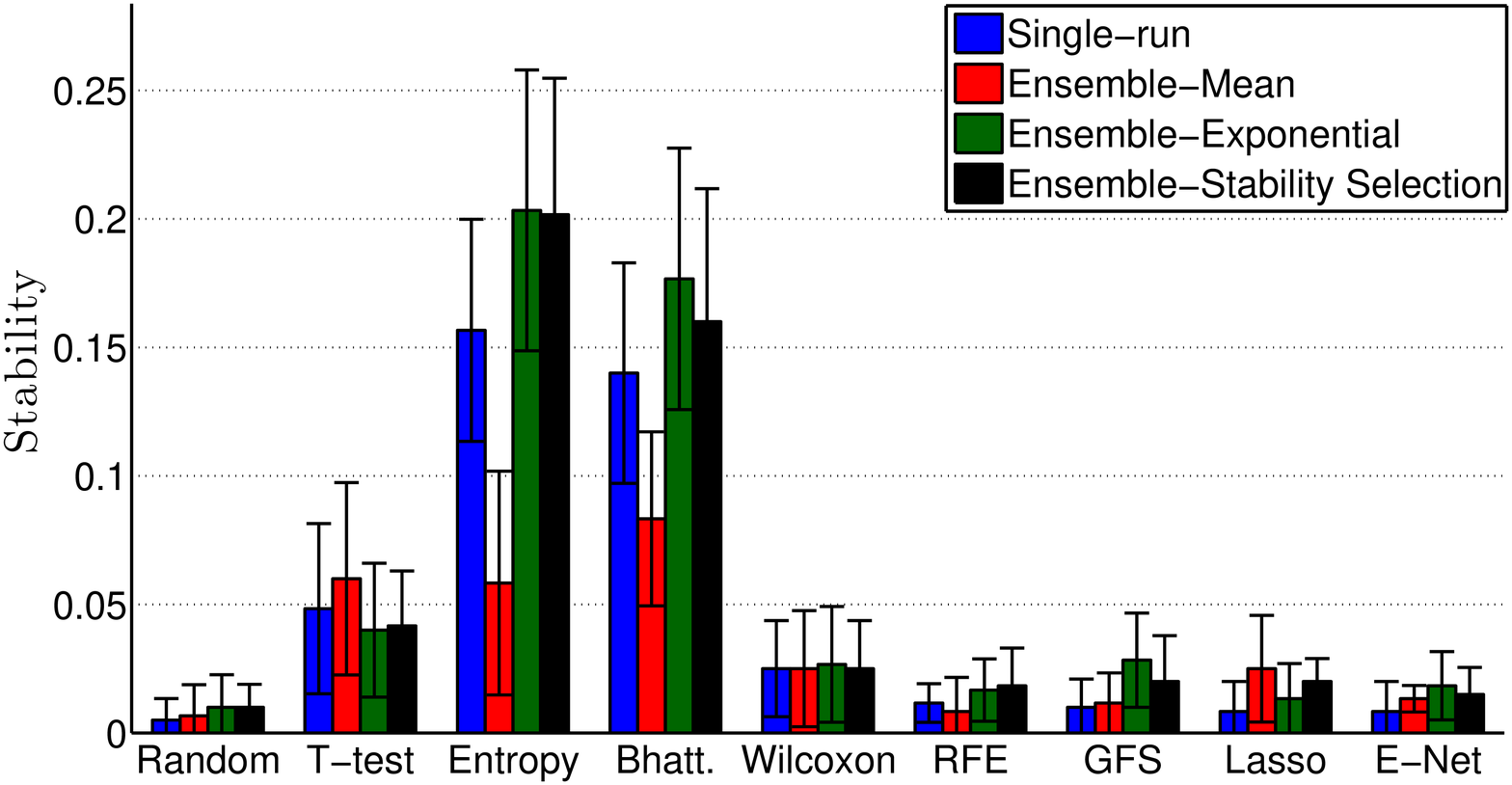}}

  \caption{Stability for a signature of size 100. Average and standard errors are obtained over the four datasets. a) Soft-perturbation setting. b) Hard-perturbation setting. c) Between-datasets setting.}
  \label{fig:stability}
\end{figure}

We now assess the stability of signatures created by different feature selection methods at the gene level. 
Figure \ref{fig:stability} compares the stability of $100$-gene signatures estimated by all feature selection methods tested in this benchmark, in the three experimental settings: soft-perturbation, hard-perturbation and between-datasets settings. The results are averaged over the bootstrap replicates and the four datasets. 
It appears very clearly and significantly that filter methods provide more stable lists than wrappers and embedded methods. It also seems that ensemble-exponential and ensemble-stability selection yield much more stable signatures than ensemble-average. It is worth noting that a significant gain in robustness through bootstrap is only observable for relative entropy and Bhattacharyya distance. Interestingly, SVM-RFE seems to benefit from ensemble aggregation in the soft-perturbation setting, as observed by \citet{Abeel2010Robust}, but this effect seems to vanish in the more relevant hard-perturbation and between-dataset settings.
We also observed that the relative stability of the different methods does not depend on the size of the signature over a wide range of values, confirming that the differences observed for signatures of size 100 reveal robust differences between the methods (Supplementary Figure 1).

Obviously, Figures \ref{fig:stab100_hard} and \ref{fig:stab_between} are very much alike while Figure \ref{fig:stab100} stands aside. They confirm that the hard-perturbation setting is the best way to estimate the behavior of the algorithms between different studies. The larger stability observed in the between-datasets setting compared to the hard perturbation setting for some methods (e.g., t-test) is essentially due to the fact that signatures are trained on more samples in the between-dataset setting, since no split is required within a dataset (Supplementary Figure 2). This suggests that, as predicted by \citet{Ein-Dor2006Thousands}, the main reason for signature instability for a given technology is really the sample size issue, and not differences in cohorts or experimental protocols.

\subsection{Interpretability and functional stability}

Even when different signatures share no or little overlap in terms of genes, it is possible that they encode the same biological processes and be useful if we can extract information about these processes from the gene lists in a robust manner. In the case of breast cancer prognostic signatures, for example, several recent studies have shown that functional analysis of the signatures can highlight coherent biological processes \citep{fan2006concordance,reyal2008comprehensive,shen2008pathway,abraham2010predicting,shi2010functional}. Just like stability at the gene level, it is therefore important to assess the stability of biological interpretation that one can extract from signatures.

First, we evaluate the \emph{interpretability} of signatures of size $100$, i.e., the ability of functional analysis to bring out a biological interpretation for a signature. 

\begin{figure}[!ht]
  \centering

 \subfloat[]{\label{fig:interpretability}\includegraphics[width=.75\textwidth]{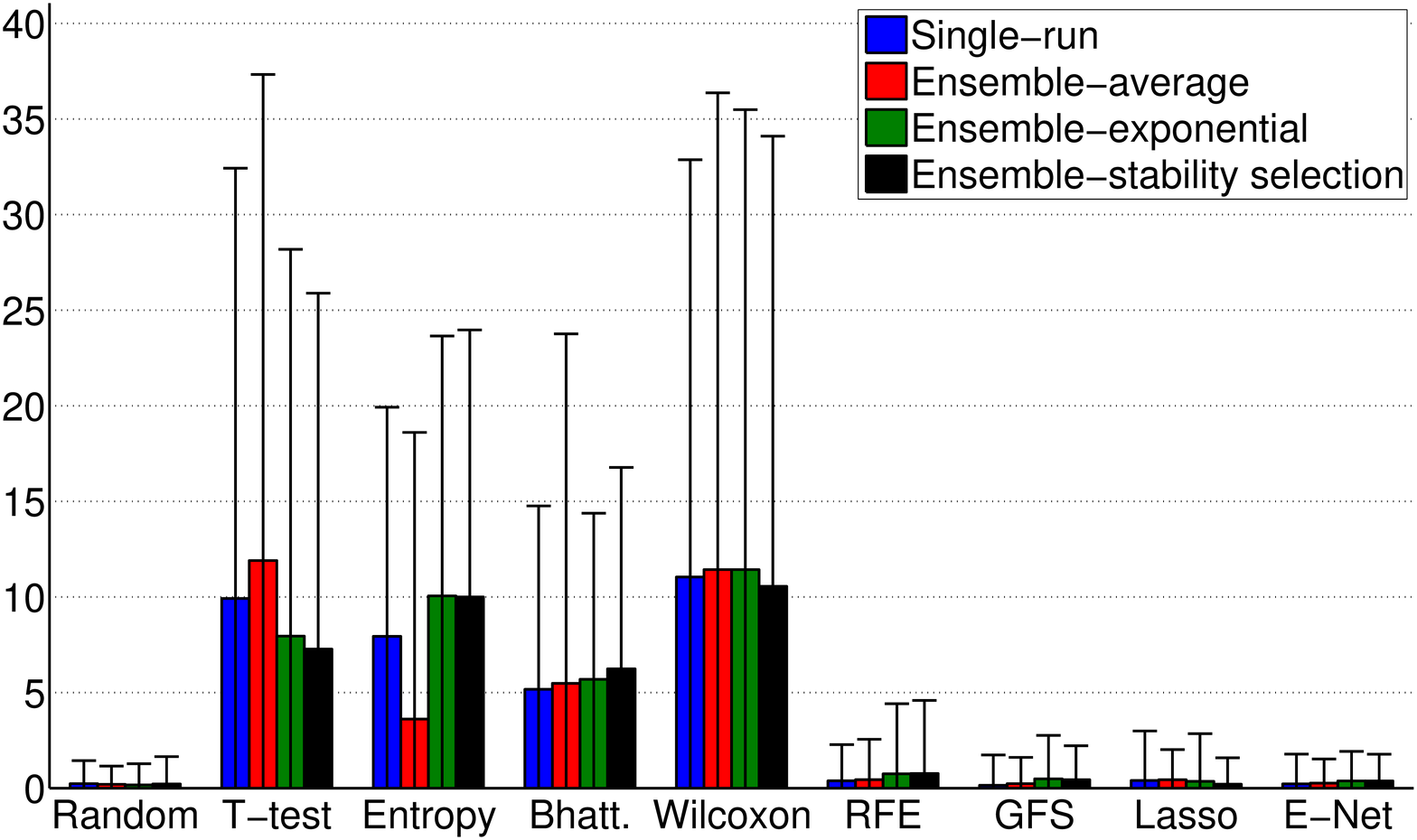}}

  \subfloat[]{\label{fig:GO_dist_inter}\includegraphics[width=.75\textwidth]{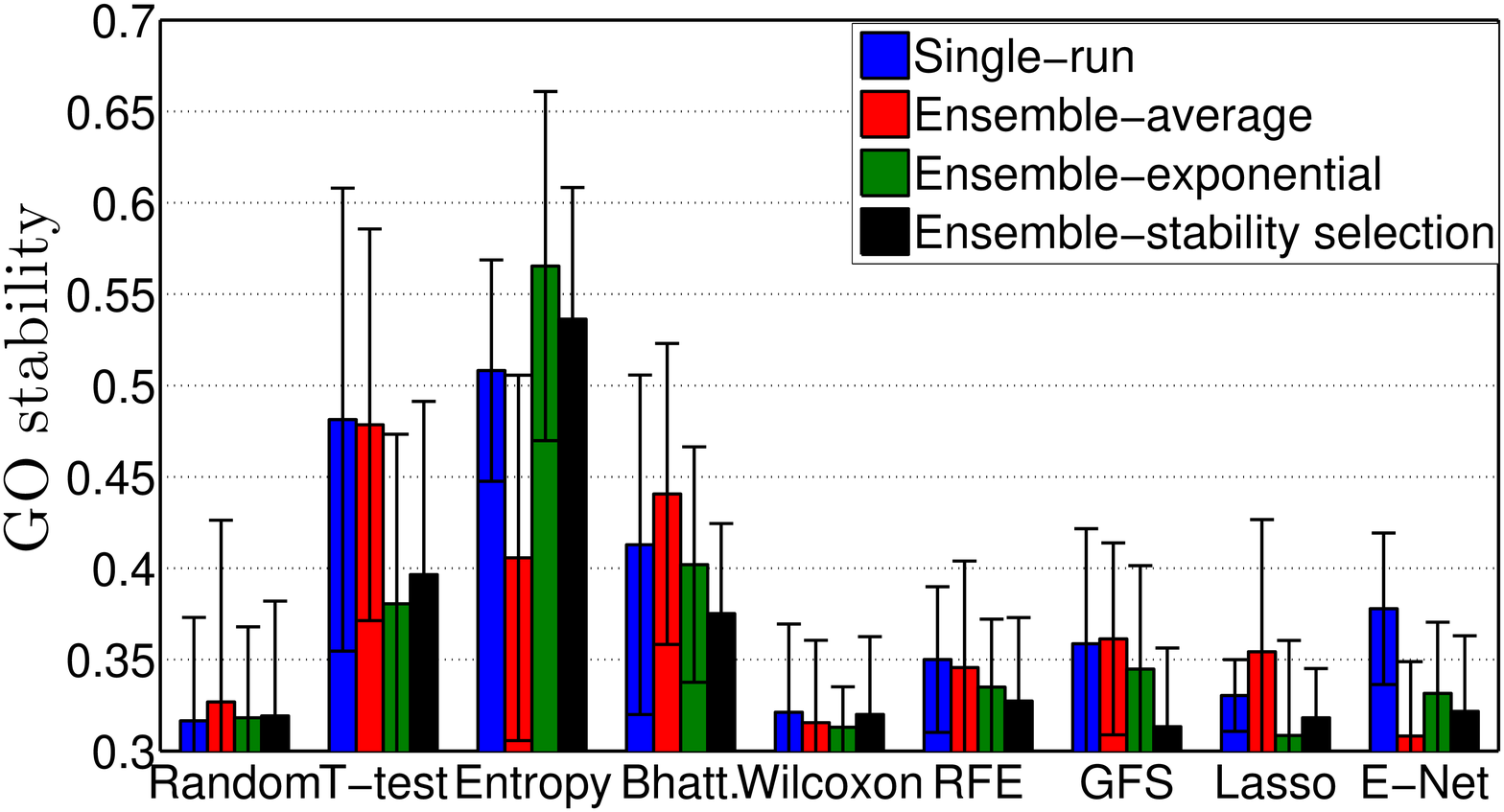}}
  \caption{GO interpretability and functional stability of for a signature of size 100. a) Average number of GO BP terms significantly over-represented. b) Functional stability in the between-datasets setting.}
  \label{fig:GO_stab}
\end{figure}

As shown on Figure \ref{fig:interpretability}, the four filter methods appear to be much more interpretable than wrappers/embedded methods. However, it should be pointed out that the number of significant GO terms is often zero regardless of the algorithm, leading to large error bars. Ensemble methods do not seem to enhance the interpretability of signatures.

Second, we assess on Figure \ref{fig:GO_dist_inter} the \emph{stability} of the functional interpretation in the between-dataset setting (the soft- and hard-perturbation settings are shown in Supplementary Figure 3).
Stability results at the functional level are overall very similar to the results at the gene level, namely, we observe that univariate filters are overall the most stable methods, and that the hard-perturbation setting returns a trustworthy estimate of the inter-datasets stability. In particular, we note that in the single-run settings, only signatures obtained from filters are significantly more stable than random at ($p<0.05)$. We also note that Ensemble-mean never improves the functional stability and that Ensemble-exponential/Ensemble stability selection return more stable signatures than single-run for Entropy and Bhattacharyya ($p<10^{-22}$) as well as for GFS ($p<10^{-6}$) and Lasso ($p=0.029$) although less significantly.

\section{Discussion}
\begin{figure*}[!ht]
\centering
\includegraphics[width=\textwidth]{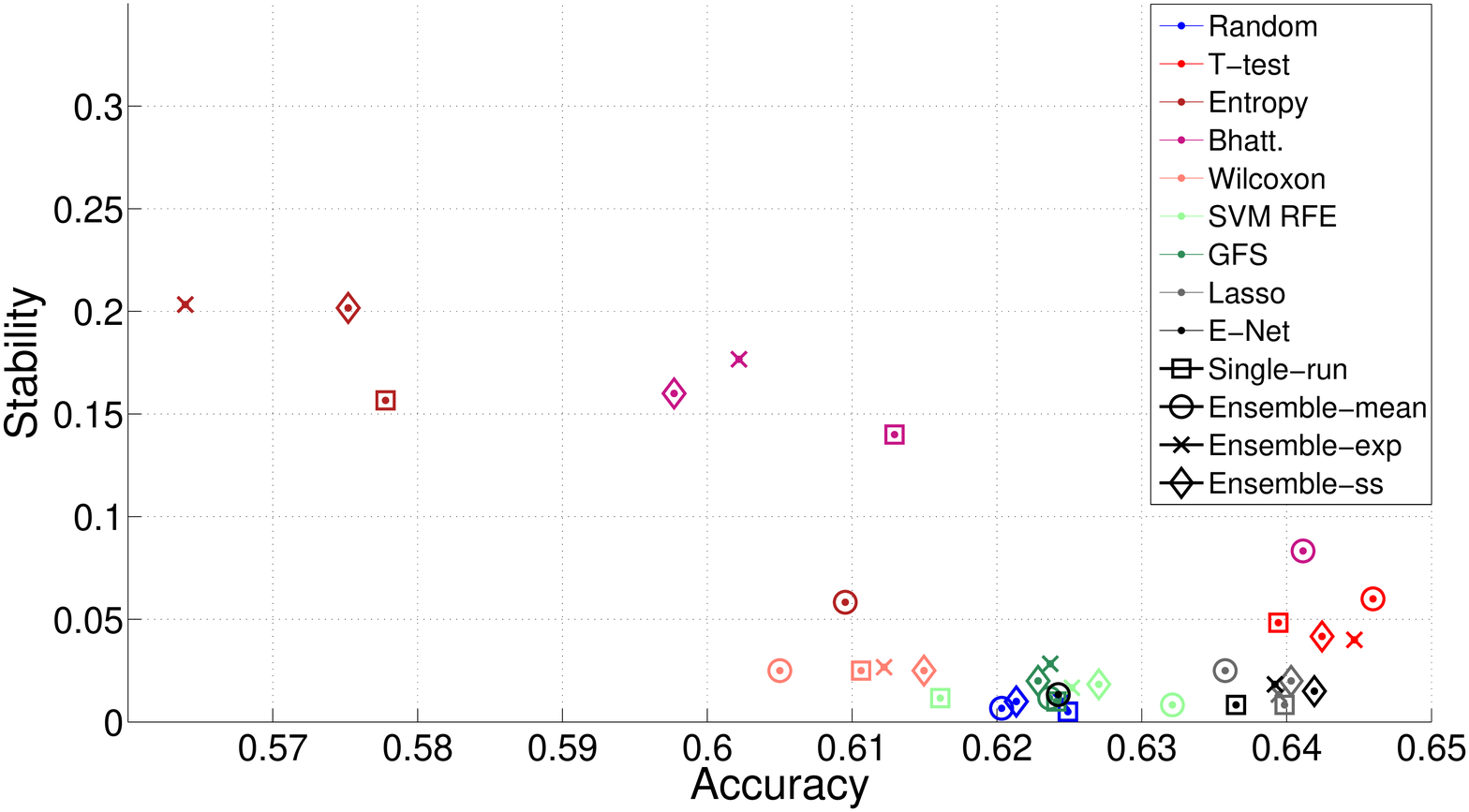}
\caption{Accuracy versus stability for each method in the between-datasets setting. We show here the average results over the four datasets.}
\label{fig:acc_stab}
\end{figure*}

We compared a panel of 32 feature selection methods in light of two important criteria: accuracy and stability, both at the gene and at the functional level. Figure \ref{fig:acc_stab} summarizes the relative performance of all methods, and deserves several comments.

Taking random feature selection as a baseline, we first notice the strange behaviour of gene selection by Batthacharyya distance and relative entropy: they are both more stable but less accurate than random selection. A careful investigation of the genes they select allowed us to identify that they tend to select genes with low expression levels, independently of the sample labels, as explained in Supplementary Figures 4 and 5. This unwanted behaviour can easily be fixed by pre-filtering genes with small variations, but it highlights the danger of blindly trusting a feature selection method, which in this case gives very stable and interpretable signatures. 

Second, we observe that among the other methods, only elastic net, Lasso and t-test clearly outperform random in terms of accuracy, and only t-test outperforms it in terms of stability.  Overall, t-test gives both the best performance and the best stability. The fact that the Lasso is not stable is not surprising since, like most multivariate methods, it tries to avoid redundant genes in a signature and should therefore not be stable in data where typically many genes encode for functionally related proteins. What was less expected is that neither the elastic net, which was designed exactly to fight this detrimental property of Lasso by allowing the selection of groups of correlated genes, nor stability selection, which is supposed to stabilize the features selected by Lasso, were significantly more stable than the Lasso. In addition, we also found very unstable behaviours at the functional level. This raises questions about the relevance of these methods for gene expression data. Similarly, the behavior of wrapper methods was overall disappointing. SVM RFE and Greedy Forward Selection are neither more accurate, nor more stable or interpretable than other methods, while their computational cost is much higher. Although we observed like \citet{Abeel2010Robust} that SVM RFE can benefit from ensemble feature selection, it remains below the t-test both in accuracy and stability.

Overall we observed that ensemble method which select features by aggregating signatures estimated on different bootstrap samples increased the stability of some methods in some cases, but did not clearly improve the best methods. Regarding the aggregation step itself, we advise against the use of ensemble-average, i.e. averaging the ranks of each gene over the bootstrapped lists, regardless of the selection method. Ensemble-stability selection or ensemble-exponential gave consistently better results. The superiority of the latter two can be explained by the high instability of the rankings, as discussed in \citet{iwamoto2010predicting}. 

Regarding the choice of method to train a classifier once features are selected, we observed that the best accuracy was achieved by the simplest one, namely the \emph{nearest centroids} classifier, used e.g. by \citet{lai2006comparison,abraham2010predicting}. An advantage of this classifier is that it does not require any parameter tuning, making the computations fast and less prone to overfitting.

We noticed that evaluating the stability and the interpretability in a soft-perturbation setting may lead to untrustworthy results. The best estimation seems to be obtained in the hard-perturbation setting experiments. The lack of stability between datasets has been explained by four arguments. First data may come from different technological platforms, which is not the case here. Second and third, there are differences in experimental protocols and in patient cohorts, which is indeed the case between datasets; fourth, the small number of sample leads statistical instability. We however obtained very similar stability in the \emph{hard-perturbation} setting (within each dataset) and in the \emph{inter-datasets} results. This suggests that the main source of instability is not the difference in cohorts or experimental protocols, but really the statistical issue of working in high dimension with few samples.

\bibliographystyle{natbib}
\bibliography{influence}

\end{document}

\end{document}